# A Multi-Server Information-Sharing Environment for Cross-Party Collaboration on A Private Cloud

Jianping Zhang, Qiang Liu, Zhenzhong Hu, Jiarui Lin*, Fangqiang Yu

Department of Civil Engineering, Tsinghua University, Beijing, China

**Abstract**

Interoperability remains the key problem in multi-discipline collaboration based on building information modeling (BIM). Although various methods have been proposed to solve the technical issues of interoperability, such as data sharing and data consistency; organizational issues, including data ownership and data privacy, remain unresolved to date. These organizational issues prevent different stakeholders from sharing their data due to concerns regarding losing control of the data. This study proposes a multi-server information-sharing approach on a private cloud after analyzing the requirements for cross-party collaboration to address the aforementioned issues and prepare for massive data handling in the near future. This approach adopts a global controller to track the location, ownership and privacy of the data, which are stored in different servers that are controlled by different parties. Furthermore, data consistency conventions, parallel sub-model extraction, and sub-model integration with model verification are investigated in depth to support information sharing in a distributed environment and to maintain data consistency. Thus, with this approach, the ownership and privacy of the data can be controlled by its owner while still enabling certain required data to be shared with other parties. Application of the multi-server approach for information interoperability and cross-party collaboration is illustrated using a real construction project of an airport terminal. Validation shows that the proposed approach is feasible for maintaining the ownership and privacy of the data while supporting cross-party data sharing and collaboration at the same time, thus avoiding possible legal problems regarding data copyrights or other legal issues.

## 1. Introduction

The architecture, engineering, construction (AEC) industry and the facilities management (FM) profession work with highly diverse sets of information and models[20], which are fragmented into different file formats and applications. An AEC/FM project requires collaboration and exchange of information among many parties, including the owners, architects, engineers, estimators, surveyors, contractors, and regulators, among others. Building information modeling (BIM) involves not only the digital representation of physical and functional characteristics of a facility but also the process of creating, using, and maintaining a shared knowledge resource that forms the basis for decision making throughout the lifecycle of a facility[6]; BIM has now emerged as the means for information exchange among the various parties involved in construction projects[15].

With years of research and development, there have been many significant developments and implementations of BIM technologies from researchers and software vendors to support information exchange and encourage collaborations among the parties involved in a project. As the AEC/FM industry moves toward the use of BIM tools, digital design models are now being embraced as the primary medium for information exchange. There is a wide variety of BIM tools that serve the AEC/FM industry, covering many different domains[6]. However, no single application can provide all the services or functionalities required by the AEC/FM industry. Because companies use different tools, each has its own internal model and data representation; thus, interoperability has become one of the main challenging issues for collaboration[15,18].

The term "interoperability" can be defined as the ability of diverse systems and organizations to work together (interoperate). There are two fundamental interoperability issues: software interoperability and organizational interoperability. Software interoperability, as noted earlier, occurs due to the diverse sets of tools used in the AEC/FM industry. Organizational interoperability occurs due to the characteristics of distributions and collaborations in different organizations. For example, a general contractor (GC) may establish a server to manage and share all the data collected from different subcontractors to form a unified model of a large project. Since the data server is controlled by the GC, subcontractor loses ownership of his data actually though he should own the data as agreed in the contract. Meanwhile, a subcontractor may just want to share his cost calculated according to the contract but not to share his real cost and work efficiency, which is his core competency. The main problem of organizational interoperability involves issues of responsibility, liability, stability, model ownership as well as data consistency, and data availability[16,29].

To date, many attempts have been made to enhance software interoperability. Among them, an open BIM data standard known as Industry Foundation Classes (IFC) has been proposed. Using the IFC standard, the BIM data are delivered through a unified and open data format to enable software tools to understand the information through the IFC schema[25]. Although most current research and development efforts focus on technical implementations to support software interoperability, organizational interoperability issues[1,30], such as who owns the model and who is responsible for the usability or the validity of the model, remain unresolved.

Previous studies have largely addressed software interoperability issues without careful consideration of organizational issues[16]. In this paper, we first provide a brief review of interoperability within the AEC/FM sector and identify some of the shortcomings of the current approaches on data exchange to support cross-party collaborative environments. To address the organizational interoperability, we propose a multi-server approach based on a private cloud platform and discuss its applicability in cross-party information exchange and sharing. Specifically, a requirement analysis of BIM data services for data exchange among stakeholders is presented. Based on the requirements analysis, a multi-server service architecture that includes technical approaches that are designed for cross-party model data distribution, integration, and management is presented. Finally, a scenario that utilizes real project data is presented to illustrate the multi-server framework.

2. **Review of current approaches on cross-party data exchange and collaboration**

This section reviews the major approaches related to organizational interoperability issues with regard to supporting cross-party and multi-disciplinary/multi-user collaborations, namely, file transfer, central database, single server and cloud-based server (see Figure 1).

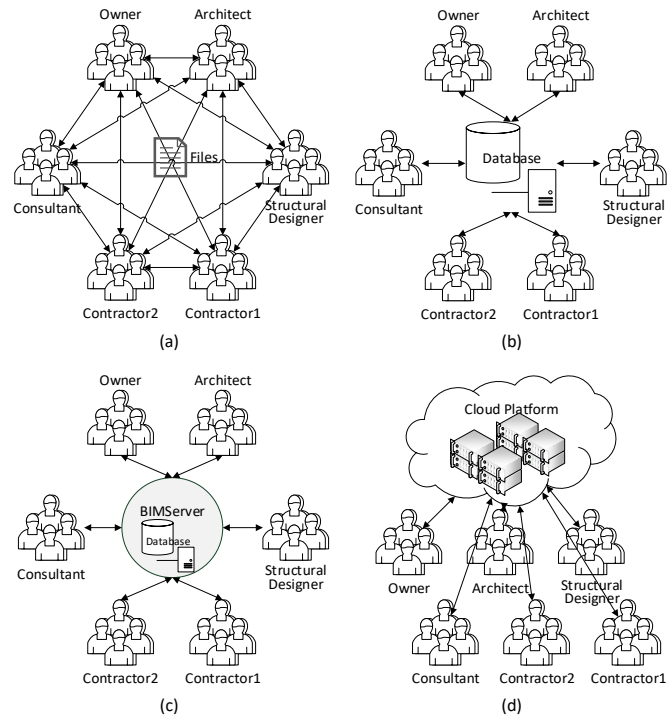

**Figure 1 BIM exchange approaches:**

**(a) file transfer, (b) central database, (c) single server, and (d) cloud server.**

2.1 File transfer

Analogous to the traditional practice of delivering data through drawings and documents prior to the popularization of computer and information technology (IT), the industry now delivers such information as electronic documents. Nevertheless, printed documents and drawings remain the most common norm for archiving and publishing project information. AEC/FM software tools often retain a proprietary internal file format. However, using STEP's EXPRESS\EXPRESS-G as the data modeling language to define entities, attributes, and constraints, the IFC standard has been formulated and designed to support model exchanges for a wide range of applications that are commonly used by the construction industry[20]. Many commercial vendors now provide IFC interfaces for their software tools. Many researchers and users have taken advantage of the IFC interface to enable software interoperability among the tools. However, even with the IFC interfaces, as illustrated in Figure 1(a), the mode of data transfer is to deliver information in the form of data files among stand-alone applications.

In a multi-party scenario with this approach, each organization has many participants, and each participant has access to many applications. This practice leads to a significant number of files being transferred back and forth during a project, and object-level management of BIM data is impossible in this approach. Thus, this approach has the inherent problems of data redundancy and inconsistency because process control is difficult to enforce.

2.2 Central database

One way to avoid the data redundancy and inconsistency problem is to employ a central data repository, such as a database. As shown in Figure 1(b), the participants of a project run their applications by accessing the data from the central, shared database. Currently, it is not uncommon that a centralized BIM database is set up by the project owner, manager, or BIM consultant for collaboration purposes. The project participants are then assigned access to the centralized BIM database. This approach provides a unified view of the BIM data for the users. However, with this approach, there lacks a generalized layer for model validation, subset extraction and integration, and thus, different participants must implement similar functions based on a central database by themselves. This approach leads to reworking the applications development and could cause data corruption and unauthorized access.

2.3 Single server

An improved method for sharing information among the participants is to set up a BIM server for different users and their systems; this approach extends beyond the central database approach and is influenced by the service-orientation concept in the software industry[28]. Different from the central database approach, the BIM server provides not only a centralized BIM database for the storage of information but also certain services (such as common functionalities and processes) for accessing information within a safe and reliable environment, as shown in Figure 1(c). Because the server takes charge of some functional tasks, the client side (e.g., BIM tools or software) can be simplified to the maximum extent, which means that lightweight clients (e.g., web browsers or mobile devices) can run BIM applications smoothly.

Specifically, aside from data, the BIM server also provides additional functions based on the data, such as model scanning, three-dimensional viewing, versioning, and conflict checking. For example, Faraj et al.[11] proposed a collaborative construction computer environment, WISPER, which was established on a web server, where the underlying data were stored using an object-oriented database designed to integrate visualization, cost estimation, project management, and supplier information. Chen et al.[23] proposed a browser-server (B/S) structured information server, in which Java and Java3D were applied to achieve user interaction and visualization; a useful algorithm to transfer the architectural model into a structural model was also proposed by analyzing the topological relationships among the building components. Plume and Mitchell[14] reported a use case of the BIM server for a collaborative design process in a teaching context where a model server based on an express data management (EDM) technology was adapted to process and share IFC model data. The report noted that the EDM model server considered overall models rather than achieving sub-model extraction or delivery. Three representative BIM server solutions are the IFC Model Server developed by VTT Building and Transport and SECOM Co. Ltd.[2], the EDM model server developed by Jotne EPM Technology[17], and the Bimserver.org developed by TNO Netherlands and TU of Eindhoven[12]. These are B/S structured or client-server (C/S) structured, and they feature the functions of importing and exporting the IFC model.

To identify what information should be shared and how to share it, BuildingSMART presented a process definition standard called Information Delivery Manual (IDM) to address when certain types of information are required during the construction of a project or the operation of a built asset[31]. BuildingSMART also presented a model view definition (MVD) to identify the prerequisites and outcomes of the processes for information exchange. An MVD defines a subset of the IFC schema that converts the IFC model into IFC sub-models according to one or many exchange requirements of a certain process[20].

This single-server approach is advantageous due to its ability to support process-oriented sub-model applications because the sub-model operations are handled on the server side. However, the model development, review, uploading, downloading, and analysis activities could be highly complex within a BIM-server environment[29]. Furthermore, it is difficult for a single server to handle massive data of a largescale project due to the limitations in the capacity for data storage and processing.

2.4 Cloud server

Accompanied by a series of technological breakthroughs, cloud computing (CC), where tools and services reside in the cloud environment and are delivered over a network, has the potential to reform the information management in the building industry. Over the past few years, CC-based BIM (or Cloud- BIM, as shown in Figure 1(d)) has emerged as a new engineering paradigm because such services enable higher performance to be achieved at a lower cost[27].

Semi-structured interview conducted by Redmond et al.[3] shows that CC can be taken as an integration platform for BIM applications and discusses the impact of CC on information exchange. With flexible data transferring and handling as well as better accessibility, CC is widely used integrated with other technologies in construction sector, for example, Fang et al.[32] integrates CC, BIM and RFID for indoor localization in construction management, Lin et al.[13] proposed a natural language based approach to intelligent BIM data retrieval and representation on cloud, while Jiao et al.[33] integrates BIM with business social networking services by adopting CC and augmented reality (AR) technology However, most of these work just utilize CC to address specific problems, but none of them focuses on the cross-party data sharing and collaboration, especially for the organizational issues. Widely used cloud BIM systems, like Autodesk BIM 360, Cadd Force, BIM9, BIMServer, BIMx and STRATUS, can be categorized into three types:

(1) cloud for virtualized desktops: this kind of system includes Cadd Force and BIM9[9].

(2) cloud for file-based BIM sharing: BIMx belongs to this category.

(3) object level BIM sharing: Autodesk BIM 360, STRUTUS and BIMServer are in this group. The first two are fixed in the Autodesk systems, while the last focuses on the support of IFC.

Unfortunately, only parts of these systems support private cloud, and only one of them support open data format like IFC[9], making it hard for stakeholders with different BIM applications to collaborate with each other. What's worse, none of them focus on the organizational issues like data ownership. In one word, many cloud servers and cloud platforms currently in use are file-based or centralized with basic object-level management of the BIM data. Cloud servers and researches regarding to organizational issues like data ownership and privacy are not too many to date. Furthermore, operating large models will lead to heavy data transfer in this scenario.

2.5 Summary and discussion

The ability of each aforementioned approach satisfies the collaboration requirements of simple projects and non-complex organizational structures. However, each approach has its own deficiencies. The central database, single-server, and cloud server approaches ensure data consistency and simplify the network topology compared to the file transfer approach. More specifically, the server-based approaches and solutions are more suitable in multi-disciplinary/multi-user scenarios due to their ability to perform sub-model operations and process control of data exchanges. Nevertheless, these three approaches confuse the users in determining where to deploy the central data repository while considering the ownership, privacy protection, and legal issues among so many participants because the data are stored centrally and the participants are normally distributed. Specifically, few participants prefer to share all the information that they generated and stored in the central data repository. Instead, every participant prefers to keep its model on its own server and share information only according to a common view, thus ensuring the ownership and privacy of its own model. However, it is difficult to define the data ownership and protect privacy in these centralized approaches because the data are stored in a central repository that is always open to all the users. It is difficult to identify who is responsible for a problem when it occurs, and specifying updating liabilities and responsibilities would require careful consideration.

## 3. Common requirements for cross-party data sharing and collaboration

Before establishing a framework for cross-party data sharing and collaboration, some common requirements should be identified. These requirements will provide important information for designing the architecture of the cloud-based multi-server approach for cross-party data exchange and collaboration. From the perspective of collaboration, common requirements that should be fulfilled are listed as follows.

### 3.1. Data ownership and privacy

Ownership of the data involves defining the individual or group of people who have certain rights and duties over the data. In the BIM process, the individual or group who owns the data typically has the right to manage it and set their authorities, and they are also responsible for the data's validity and the consequences of using it.

Privacy of the data refers to the ability of an individual or group to seclude the sensitive data, thereby protecting it from being operated on or revealed by others. In the BIM data-sharing processes, privacy can be ensured by the authority's control. Participants should have the ability to set up their own servers within the environment and control the accessibility of the data. In this manner, data privacy can be significantly increased compared with centralized approaches.

### 3.2. Data distribution

From the perspective of building lifecycle information management, the BIM data have the features of dynamic creation and natural distribution[21]. Provided that each party involved in the project owns their own data, the distribution feature of the BIM data is emphasized. Specifically, the creation of BIM data depends highly on the delivery method and the progress of the project. Thus, the environment should be flexible, which makes it possible to incorporate more stakeholders dynamically and manage more data by extension and horizontal scale. Additionally, when participants finish their tasks and then leave the project, the platform should securely transfer and deliver their data to other related parties.

### 3.3. Model integration

In construction projects, multiple parties must collaborate on common goals, such as scheduling, quality control, and safety. To ensure accuracy and readability, the information exchange should follow some common data standards and protocols, such as the IFC. Then, the exchanged information should be integrated with existing data while following a logical structure, such as a data model[29]. Therefore, the distributed information in the multi-server environment requires common management[21], which involves not only the need for recording and tracking the process but also the need to establish a global-level model (for the entire project) to support integrated applications.

### 3.4. Sub-model support

A BIM application typically involves a single or few business processes, which require only a subset of the project data; for example, RFID-based indoor localization needs geometric information in a BIM model for sensor planning[22] while cost estimation requires the quantity of different elements[24] and detailed construction scheduling requires information related to activities, resources, etc.[10] in a BIM model. Thus, it is not necessary to exchange the model that contains all the data[29]. To exchange data for these business processes, we typically include many "acquisition–processing–return" processes, and the environment should support the integration and extraction of sub-models. These requirements are more complex in the multi-server environment than in other environments because the data sources are distributed in different locations and organizations.

### 4. Cloud-based multi-server approach for cross-party data sharing and collaboration

This paper proposes a cloud-based multi-server approach for cross-party data exchange and sharing to address the above-mentioned problems and fulfill the requirements for cross-party data sharing, as shown in Figure 2. This approach allows the parties involved to establish their own BIM servers or private cloud based on Hadoop and its corresponding technologies for data sharing and collaboration in their own organizations, thus ensuring data privacy and ownership. At the same time, a global controller is proposed to register and track the location and authorization of the data, forming a unified private cloud to connect the BIM servers of the different stakeholders. In this manner, users can access private data through their own servers and the shared data from other parties by a global controller. In other words, this approach enables individuals to work with one another both in- and cross-party.

To facilitate data sharing and form a useful model, MVD is used to define which part of the model a party want share. To ensure the usefulness of the shared data, definition process of MVD should follow existing method[8] and the defined MVDs should be approved by project leader or agreed by multiple stakeholders. Therefore, since each party of the project shares necessary data to the others, it is possible to form a useful model by integrating shared data of different parties.

As illustrated in Figure 2, the global controller provides functions for shared data registration and indexing, cross-party authorization, party server management, model view management and model integration. At the same time, the server of each party is responsible for in-party authorization and server management, shared data caching or replication persistence, and model extraction/integration/validation.

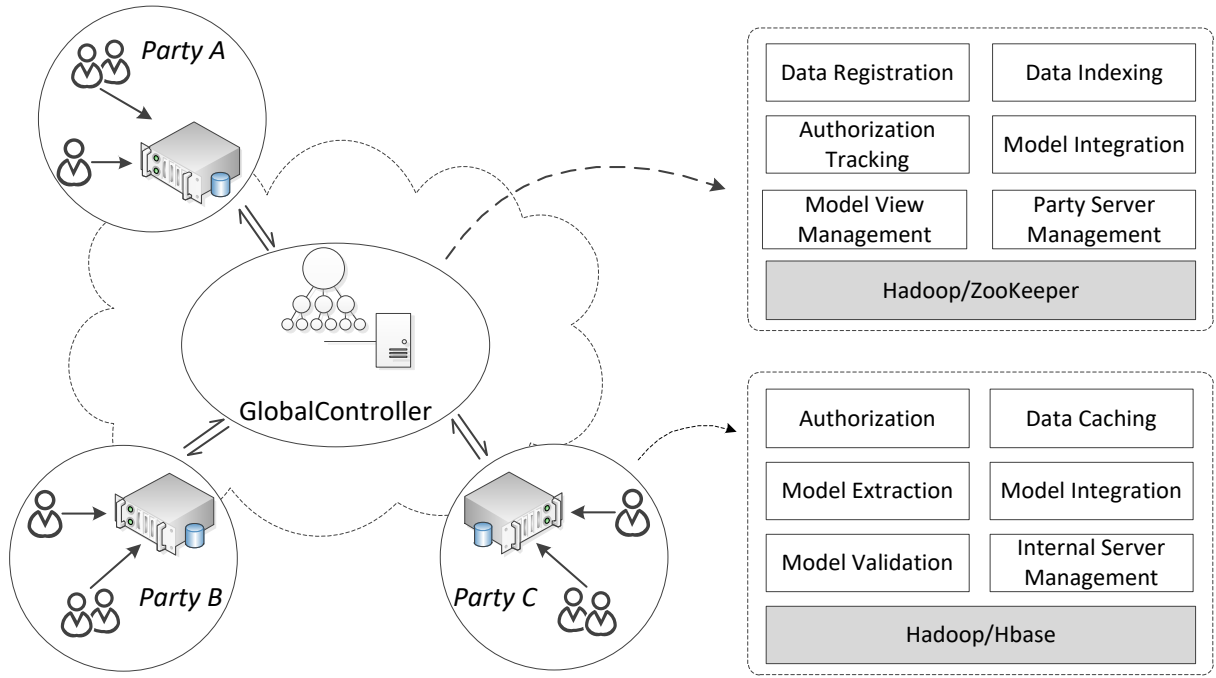

**Figure 2 Multi-server approach for cross-party data sharing and collaboration**

The proposed approach is established on the most popular framework – Hadoop – to create a cloud, that is to say, Hbase is adopted as underlying data storage for the proposed approach, and Zookeeper is utilized in global controller for global data registration and multi-server tracking. In this manner, basic cloud features like horizontal scaling, big data handling are already built in Hadoop, thus reducing the work to establish a distributed system from scratch. In addition, with data versioning feature ready in Hbase, it is easy to persist multiple versions and track the changes of the BIM model. In other words, a structural engineer can easily see the changes between consecutive architecture models and find possible impacts of these changes on his work.

In this approach, the data owned by different stakeholders was kept in different locations and had different permissions. These features will increase the complexity involved in maintaining the consistency and integrity of the BIM model. In other words, it is difficult to track the data changes, extract the required sub-models and integrate the sub-models from different parties to support a specific application. Thus, the following parts of this section will focus on three parts: global data controlling and authorization management, distributed sub-model extraction and sub-model integration based on MVD.

### 4.1. Global data control and authorization management

**1) Requirement-driven data distribution**

In the above-mentioned environment, the server of each participant stores not only the self-produced data but also the necessary data produced by other participants because of the following considerations:

a) The environment should provide users with the ability to derive their required data, including their own data and the shared data from other participants.

b) Users should access their own server to obtain the data that they need first, thus maximizing the bandwidth advantages of the LAN connections.

c) Unnecessary data should not be stored locally, thereby saving storage resources to improve the

efficiency of the data queries in local databases.

Requirement-driven distribution is a storage mode characterized by resource balancing and efficiency optimization. To implement this setting, the requirements should be defined first. Various existing delivery methods for construction projects prove that the data requirements of the participants are highly diverse and are not static as the project progresses. Therefore, a fast and flexible way to define the requirements is needed[8].

The MVD is widely used to define the exchange requirements in the IDM process. It is also suitable for defining the data requirements of the parties involved in a particular project. The difference is that the requirement for a party does not face a single business process but expresses the total demands for the exchange requirements of all involved business processes within a party. In the MVD-based distribution shown in Figure 3, the party server stores data selectively according to their requirements (or MVDs). Once the requirements are defined, the shared data should be automatically integrated into the database on the party server in accordance with the scope and constraints described in the MVDs.

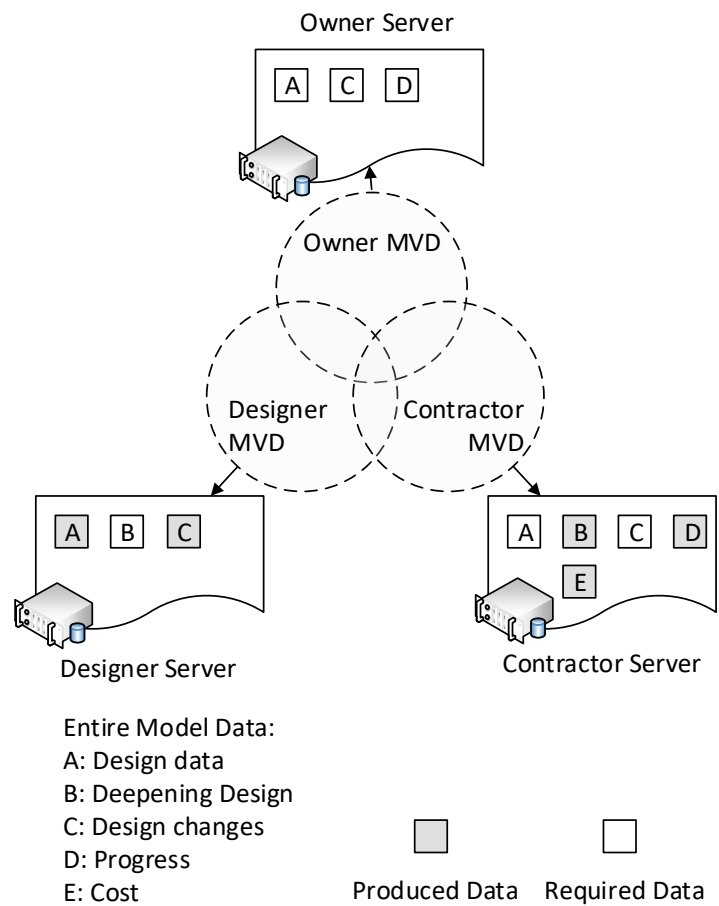

**Figure 3 Schematic view of requirement-driven data distribution based on MVD**

2) Cloud controller and global data control

The requirement-driven data distribution pattern leads the data to their destinations according to the needs of different parties, thereby forming a distribution in the multi-server data-sharing environment. Another important issue is how to manage the distributed data, which involves data query, update, and consistency maintenance. In this study, a server for the entire project, called a cloud controller, is designed to support the management of the distributed servers and data in the multi-server data-sharing environment. The core service of the cloud controller is an indexing service for recording the distributed data and marking their status, including ownership and permissions.

The main purpose of the indexing service provided by the cloud controller is to ensure the availability of the distributed data to all the parties. Major functions of the indexing service include recording and locating data objects and responding to query requests related to data appending, updating, transferring, and other processes. The indexing service is designed to target the individual objects because the in-depth collaboration in BIM processes requires data management at the object level[5,29]. The principle behind object-level data indexing is to establish and maintain ternary relations among the data objects, servers, and data access as well as to support the multi-dimensional querying of these relations. The metadata are used to build this indexing service to avoid storing large volumes of data on the cloud controller. The framework of the indexing service and the structure of the metadata are shown in Figure 4.

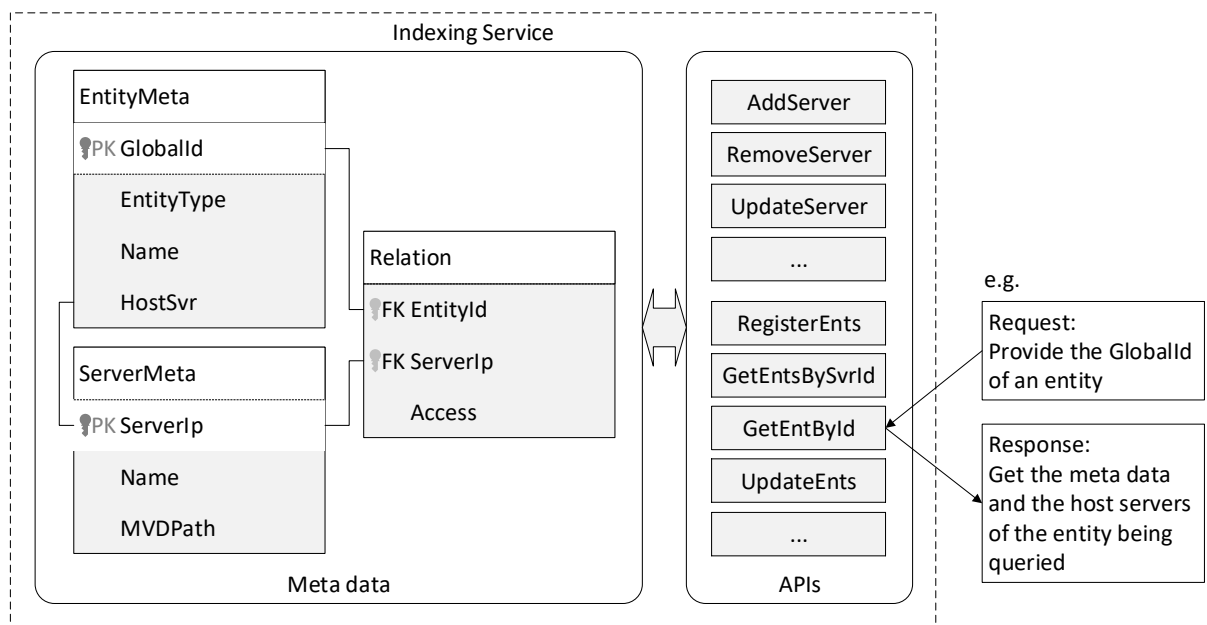

**Figure 4 Data model for global data control**

### 3) Consistency maintenance

According to the proposed distribution mode, the party servers store the concerned data as described by their MVDs. A problem is that the party's MVDs could overlap because some data are required by more than one participant, such as the building elements and spatial structure. Such overlapping in the environment inevitably results in redundant data, which means that several copies of one data object could exist.

The benefit of such redundancy is that the availability of the data is greatly increased; however, it brings a consistency issue between the different data copies. The indexing service discussed in the previous section can build indexes for all the copies of entities such that it supports the data comparison and synchronization required for consistency maintenance. Several general conventions are proposed with consideration of the engineering practices to simplify the comparison and synchronization process:

a) In the initial state, the producer is the owner of the data. (The terms "producer" and "owner" refer to different parties in a project.)

b) A data object has one and only one owner at a given point in time.

c) Only the owner has "write" access to the data.

d) The ownership of the data can be transferred to another party.

These conventions ensure the uniqueness of the source for each data object (exchangeable IFC entity) in the multi-server environment. As shown in Figure 4, the relationship between the host server and entity is "1: n" in the indexing such that only one party owns a data object at a time, which corresponds to the conventions. With these conventions, consistency of the BIM data can be achieved by controlling the data access of the parties and establishing a change propagation mechanism. These conventions also link the consistency to ownership, which simplifies the consistency issue of distributed systems.

**4) Authorization management**

According to the above-mentioned conventions for data consistency, the right for data editing was tied to the ownership of the data, which means that only the owner can change or update the data and other users have only the right to view the data. Therefore, data that persisted in the server of one party can be categorized by ownership into internal data owned by the party and external data shared by others (Figure 5). Thus, data owned by the party can be read or written by the people of the party, and whether the people of the party have the right to change the data or not is controlled by the manager of the party. Furthermore, the manager of one party can define which part of the data can be shared with other parties, and other parties have no access to unshared data by the party; thus, the privacy of the data is fulfilled. Metadata of those shared data are registered in the global controller, and other parties can obtain the metadata to create copies or replications of entities in the shared data according to the information requirements defined by the MVD. As noted in the data consistency conventions, people of the party have no right to edit or change copies or make replications of the shared data.

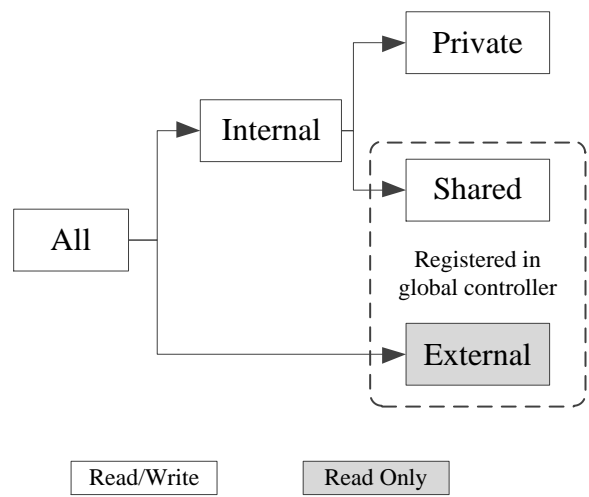

**Figure 5 Local control of authorization**

To control the authorization of the data, a property for the data access level was defined when designing the BIM database, and value of the property can be chosen from *Private*, *Shared* or *External*. For the global controller, the shared data of all parties forms a basic shared model of the project, and the external data of each party is redundant and keeps the replications of the shared data. The global controller has a service to index the location of the data and its replications, and once the source of the data is changed, its replications will be updated automatically.

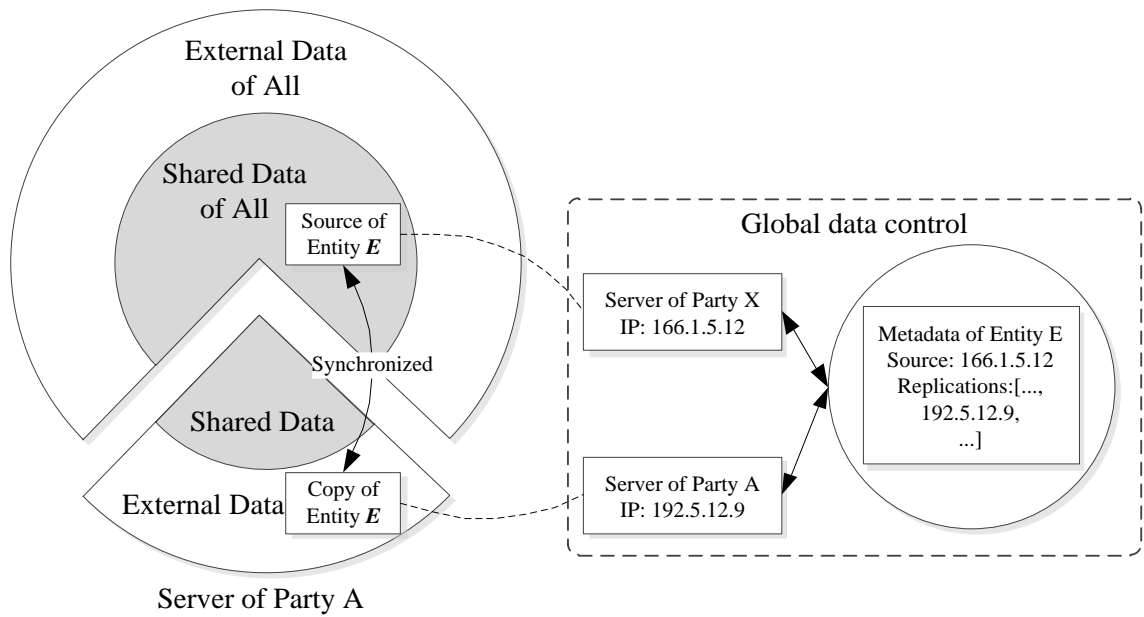

**Figure 6 Global control of authorization**

### 4.2. Distributed sub-model extraction based on the MVD

Sub-model extraction is an essential component for data interoperability and multi-stakeholder collaboration. Previous research sub-model extraction proposed different data extraction methods based on the grammar of EXPRESS, with predefined rules. Recently, BuildingSMART proposed an XML-based format MVDXML to encode an MVD that corresponds to an exchange requirement defined by an IDM process. An MVDXML file defines the allowable values at particular attributes of particular data types. In simple cases, such rules could define a single attribute on a single data type, whereas a more complex case could consist of graphs of objects and collections[26]. MVDXML can be used statically to support a specific model view or dynamically to support any model view, and thus, MVDXML can be used to filter the required data within a model view from files or servers and to validate the data to ensure that it contains the required information. To help in data filtering or extraction, MVDXML provides *Rule* (including *EntityRule* and *AttributeRule*) and *Constraint* to define the conditions that the data type, data property or data value should fulfill. With *Rule* and *Constraint*, *ConceptTemplate* and *ExchangeRequirement* can be specified and defined, and finally, a model view is established by composing or referring to *Concept* and *ExchangeRequirement*. In other words, a model view is a graph that chains all *Rule* and *Constraint* together.

**1) Process for sub-model extraction**

The basic process for sub-model extraction, according to the definition and structure of the BIM model, is shown in Figure 7. The first step to start the sub-model extraction is reading or parsing the model view defined in MVDXML; next, the graph structure of a model view is traversed to obtain all the rules and their corresponding constraints; then, all the entities in the model are iterated and checked against the rules and constraints to filter out the required entities; and finally, a sub-model is created based on the extracted entities.

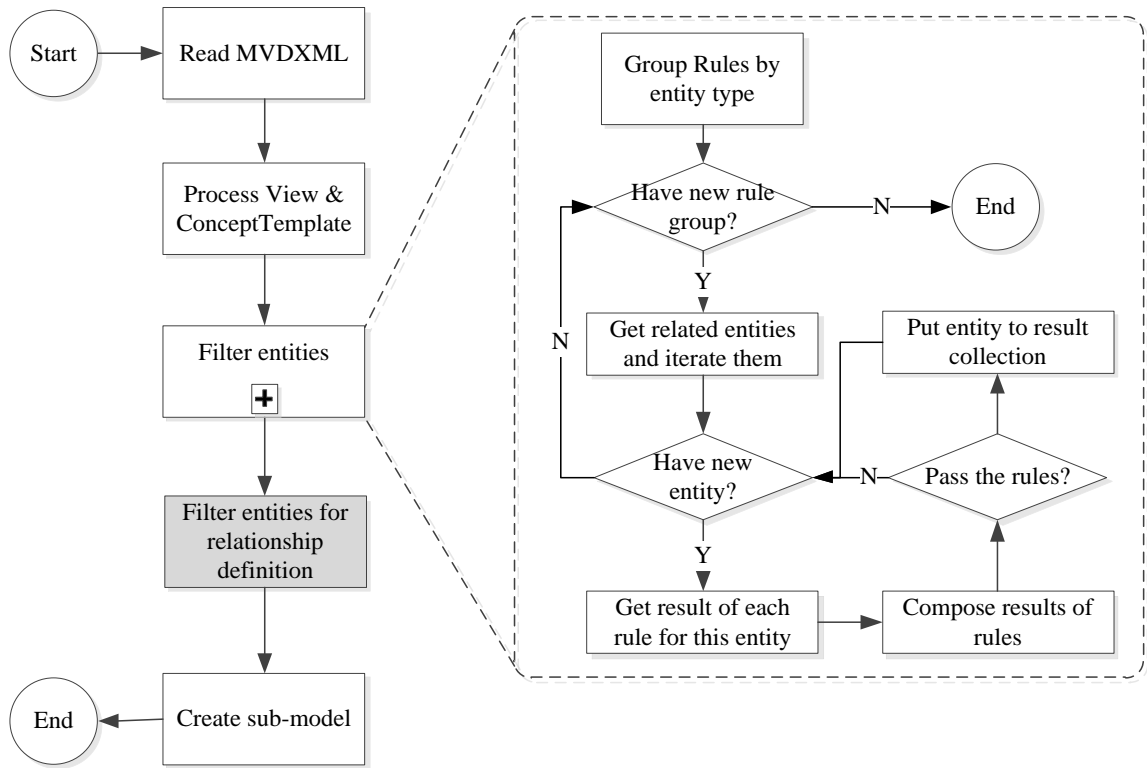

**Figure 7 Basic process for sub-model extraction**

However, this type of model extraction process could cause some problems when handling the entities that define the relationship between different entities. Taking Figure 8 as an example, the original model is composed of 2 *IfcTask* (one with the name "Zone A" and the other with the name "Zone B"), 2 *IfcColumn*, 1 *IfcBeam* and 1 *IfcSlab*. If a model view that wants to extract *IfcTask* with name equals "Zone B" and *IfcColumn* as at left bottom of Figure 8 is provided, then different results could occur. One understanding is taking all *IfcColumn* and *IfcTask* with the name "Zone B" out and omitting the relationship between *IfcTask* and *IfcColumn*. Another understanding of the model view is simply extracting *IfcTask* with the name "Zone B" out and its related *IfcColumn*. The latter approach performs better than the former approach in terms of model integrity and data privacy. Thus, a new step to extract entities for a relationship definition is added in Figure 7.

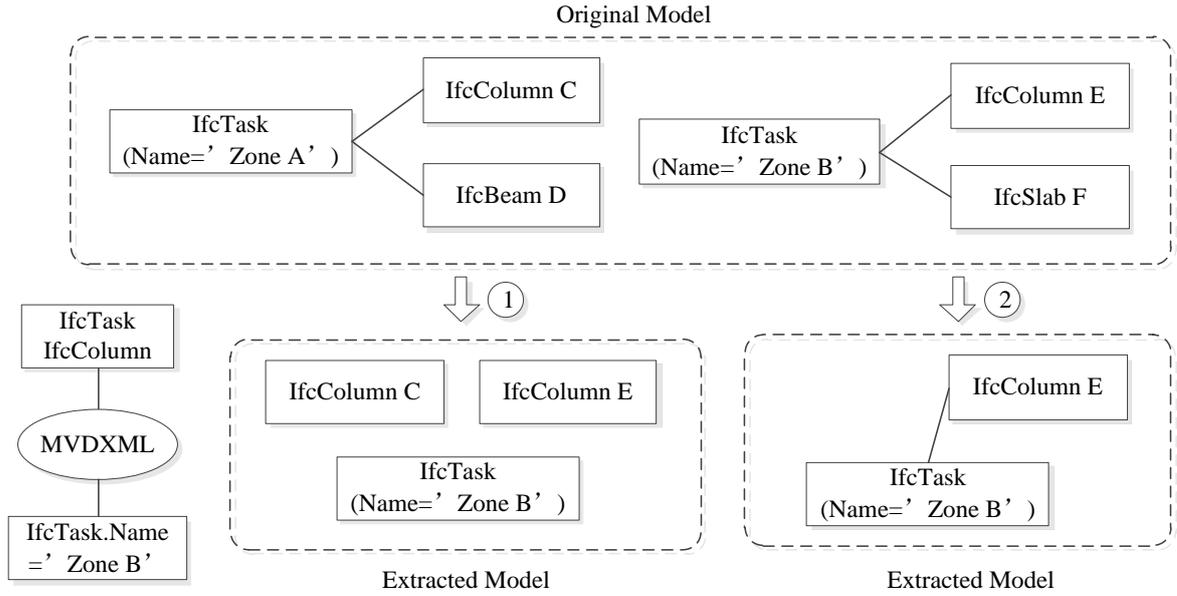

**Figure 8 Example for problems in the sub-model extraction process**

In fact, all the entities for relationship definitions in IFC are inherited from *IfcRelationship*, and the relationship between different entities is defined by the properties with names that start with "Related" or "Relating" of *IfcRelationship*. Thus, the process behind relationship entity extraction is straightforward. First, obtain all the entities inherited from *IfcRelationship* and check the properties whose name has "Related" or "Relating" as the prefix. If a property that starts with "Related" and a property that starts with "Relating" have related entities in the entity collection that was extracted in the previous steps, then this entity should be extracted and added to the final entity collection for a relationship definition. Otherwise, skip the relational entity, and continue to the next.

**2) Parallel and distributed data extraction process**

Currently, an increasing amount of data is being collected and integrated into BIM models, and companies must improve their ability to handle largescale data sets in real time. This trend is also the reason why cloud computing was proposed. However, in the above-mentioned data extraction process, all the steps are in sequential order, and the advantage of parallel data processing is not fully utilized. Because there are different replications of the data, the data extraction or filter process can be executed on different data nodes in parallel, thus increasing the data processing speed. In this research, three types of methods are proposed to parallelize and distribute the data extraction process: server-level parallelism, type-level parallelism, and instance-level parallelism (Figure 9). Shared data of one party is copied to the servers of the other parties; in other words, we have many replications of the data, and thus, the efficiency can be increased by executing the filtering or extraction process among different servers. Furthermore, the rules defined in a MVDXML can be grouped into different parts by their related data types, and each part has no relation to the others; thus, it is possible to check the rules for different data types in parallel. Finally, for entity instances of a specific data type, any combination of the rules can be rewritten as follows:

$$C = C_{b1} \cup C_{b2} \cup \cdots \cup C_{bn}$$

where

$$C_{bi} = C_{s1} \cap C_{s2} \cap \cdots \cap C_{sm}$$

In the above equations, $C_x$ denotes a single rule or a combination of rules. Additionally, $C_{xa} \cap C_{xb}$ is fulfilled only if both $C_{xa}$ and $C_{xb}$ are fulfilled, whereas $C_{xa} \cup C_{xb}$ is fulfilled if at least one of $C_{xa}$ and $C_{xb}$ is fulfilled. thus, the calculation or checking of $C_{bi}$ can be executed in parallel, thus improving the overall process slightly.

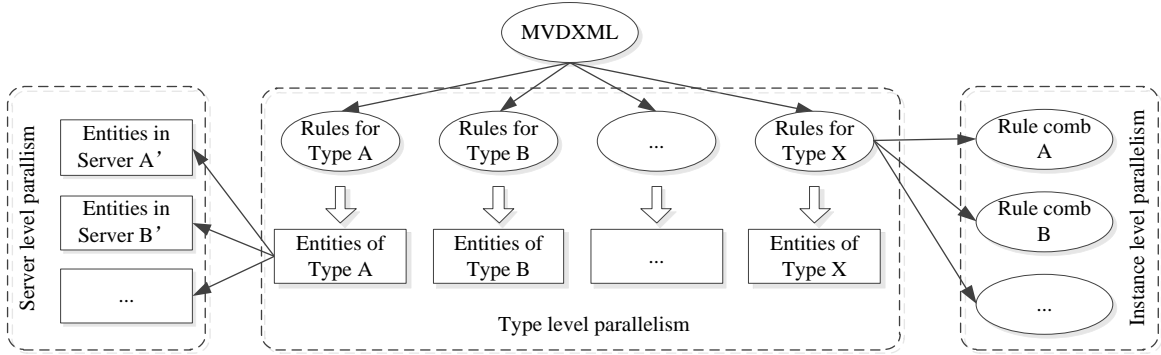

**Figure 9 Parallelizing the model extraction process at different levels**

### 4.3. Sub-model integration

Sub-model integration involves integrating one model data set into another and form a new set, where all three data sets should be self-contained and can thus be considered sub-models. More specifically, all the references and relationships in a sub-model should make sense (no null pointers, and all must be logically correct). In this approach, with the sub-model integration method, the sub-models distributed in multiple parties can be unified as a whole and the user end applications based on a small size sub-model data set can work.

**1) Exchangeable entity**

The IFC entities can be divided into two categories. 1) The rooted entities are those that are inherited from *IfcRoot* and have the *GlobalId* attribute, and thus, they are uniquely identified throughout the data exchange process. 2) The resource entities are IFC entities other than rooted entities, and they cannot exist in the data model independently and will always be referenced by the rooted entities.

Although there is no explicit restriction on the number of references of a resource entity in the IFC, we cannot assume that the two rooted entities that refer to the same resource entity always have precisely the same attribute value (when one of them changes, the other changes). Therefore, we define that a rooted entity and all the resource entities referenced by its attributes constitute an exchangeable entity. All statements about entities are about exchangeable entities in the following sections.

**2) Entity state analysis**

In a typical extract-modify-integrate process, the entities in the modified sub-model should have three states: ADD, UPDATE and DELETE. It is relatively straightforward to distinguish the three states of the entities when comparing the integrating sub-model and original extracted sub-model. However, that requires us to record all the extracted sub-models, and the sub-models not extracted from the model cannot be integrated correctly (all the entities recognized as the ADD state). This paper presents a flexible method to analyze the entity states, and the ADD and UPDATE states are assessed based on the existence of the entity (exist: UPDATE; not exist: ADD). The difficulty lies in the judgement of DELETE state. The method considers the DELETE state as a mark on the entity but does not actually remove the entity from the sub-model; the DELETE mark is expressed using the *ChangeAction* attribute of *IfcOwnerHistory*.

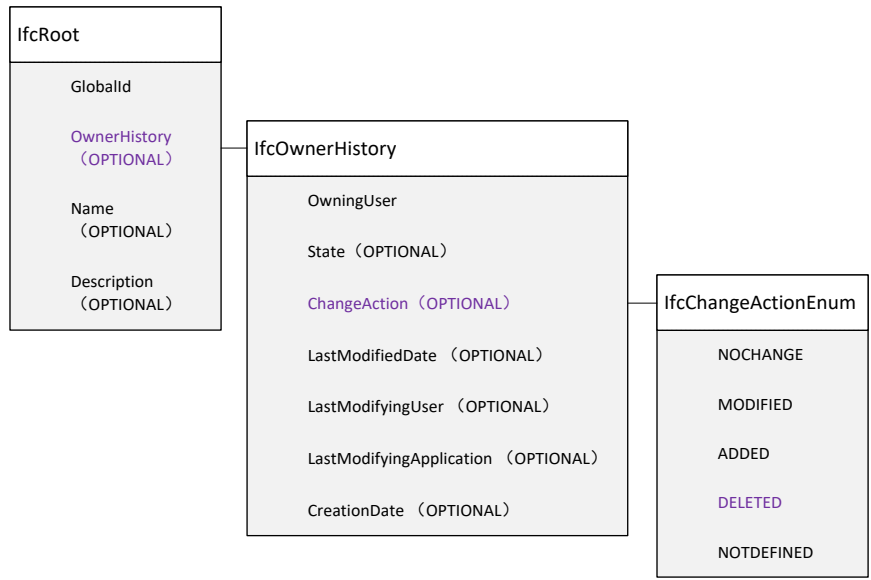

**Figure 10 DELETE mark representation in the IFC**

Because *OwnerHistory* is an optional attribute of *IfcRoot* and *ChangeAction* is an optional attribute of *IfcOwnerHistory*, it is possible that the entities deleted in the sub-model do not have the DELETE mark, and thus, they will not be deleted from the integrated model. From another perspective, the entities in the data models are related, and the following section presents a method to correct potential relationship errors during the integration, including assessing the DELETE relationships.

**3) Relationship correction**

In the IFC, one rooted entity relates to another through relation entities (inherited from *IfcRelationship*), and there are two types of relationships: one-to-one [1:1] and one-to-many [1:n]. The relationship correction is for logical errors that could arise based on the previous analysis. Because the entities with the ADD state do not affect the correctness of the relationships in the integrated model, situations in which the relationship already exists are within the scope of consideration.

This paper lists the integrating situations of [1:1] relationships and [1:n] relationships separately and proposes reasonable integration results for each situation, from which we can discover what corrections must be made based on the previous state analysis.

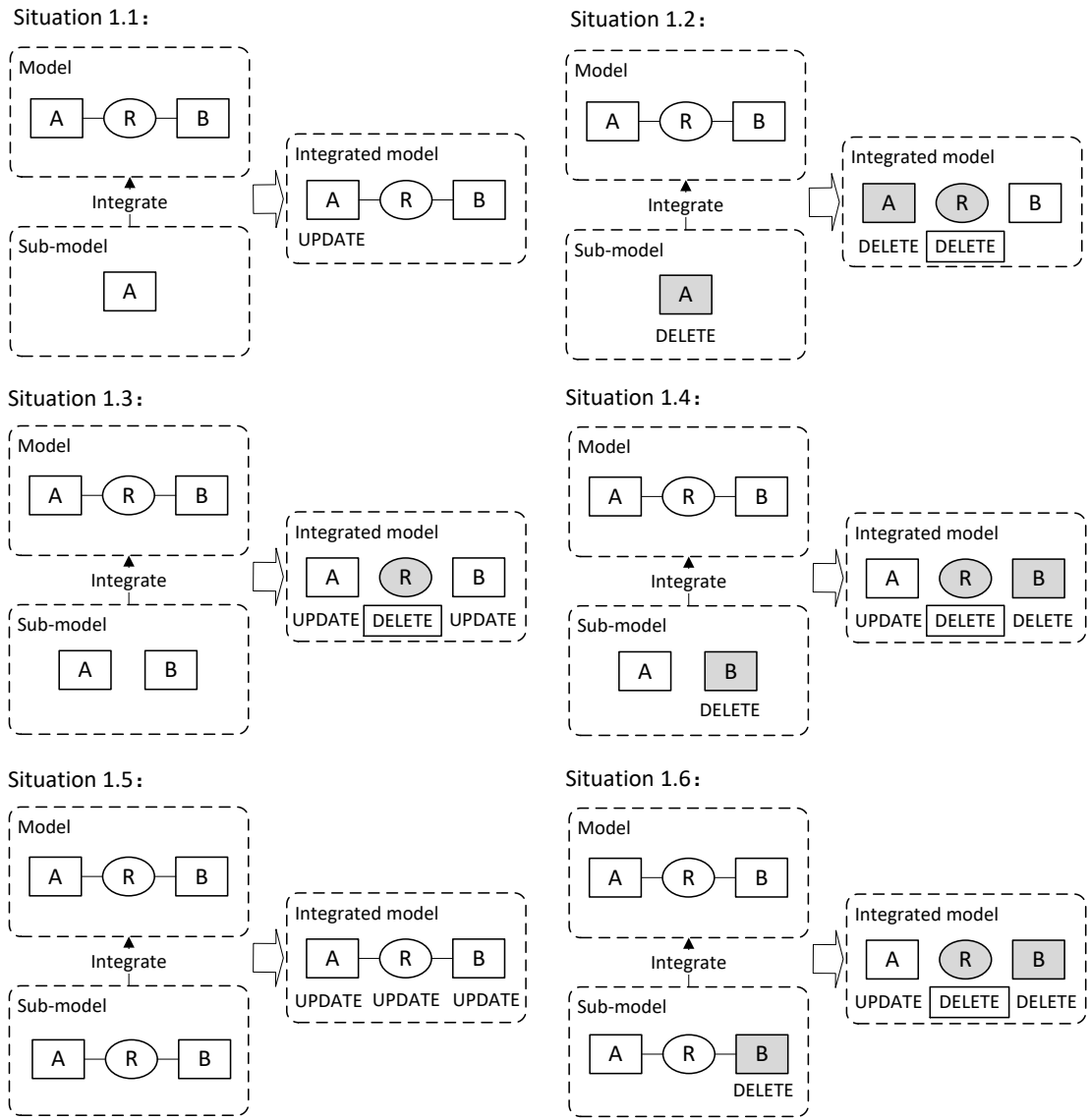

**Figure 11 Integrating situations of [1:1] relationships**

For the [1:1] situations, the "A-R-B" relationship already exists in the data model. A and B are rooted entities, and R is a [1:1] relation entity. Situations 1.1 and 1.2: only entity A is in the sub-model, and if A is marked DELETE, then the relation entity R must be deleted; Situation 1.3 and 1.4: entities A and B are both in the sub-model, but the relation entity R is not. It is reasonable to delete R in the integrated model; Situation 1.5 and 1.6: entities A, B and R are all in the sub-model. If B is marked DELETE, then R must be deleted.

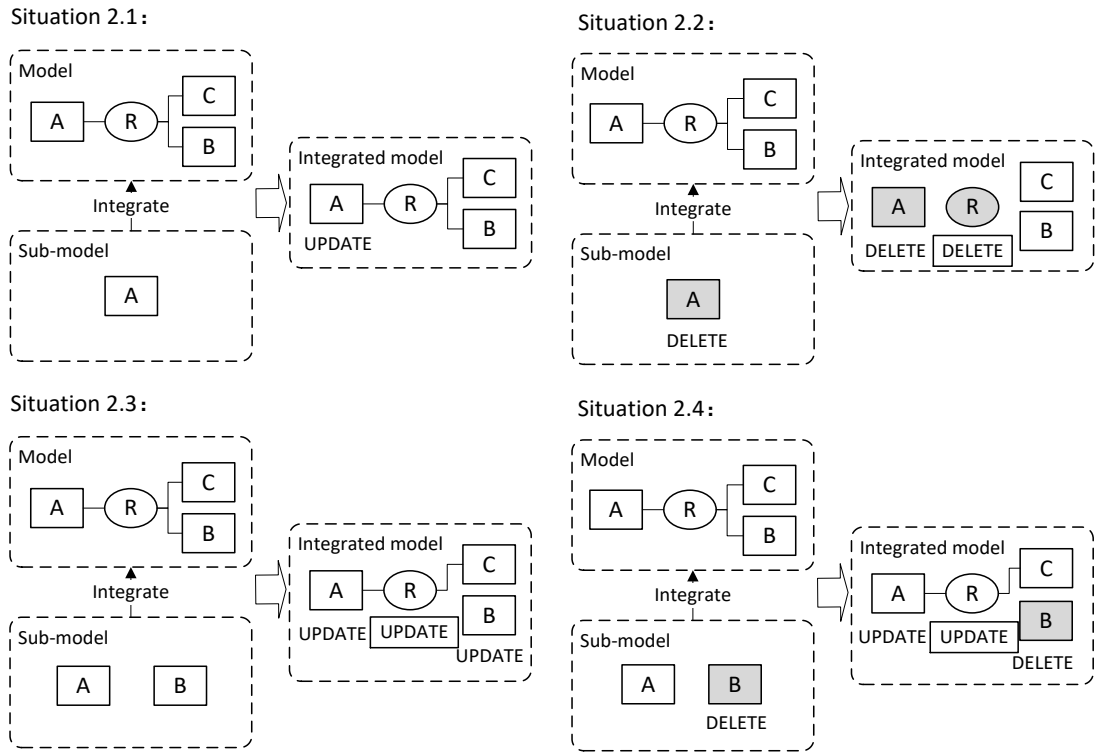

**Figure 12 Integrating situations 2.1-2.4 of [1:n] relationships**

For the [1:n] situations, the "A-R-[B, C]" relationship already exists in the data model. A is the relating entity, B and C are the related entities, and R is a [1:n] relation entity. Situations 2.1 and 2.2: only the relating entity A is in the sub-model. If A is marked DELETE, then the relation entity R must be deleted. Situations 2.3 and 2.4: the relating entity A and a relating entity B (not specific) are in the sub-model, and the relation R is not. Given that A is marked DELETE. Regardless of whether B is marked DELETE or not, B must be removed from the related entities list from R (update R).

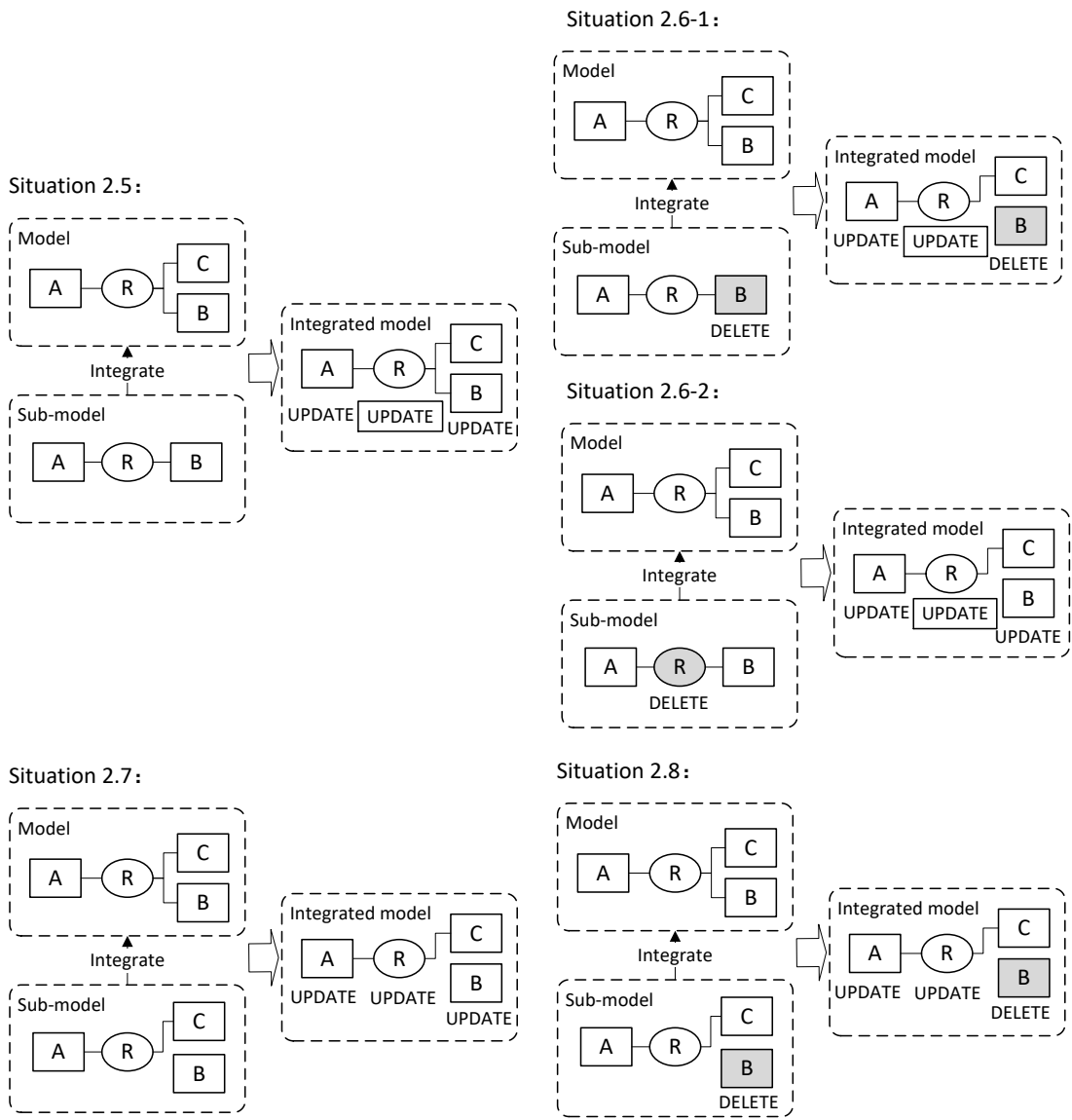

**Figure 13 Integrating situations 2.5-2.8 of [1:n] relationships**

Situation 2.5: no entity needs to DELETE, update all entities. Situation 2.6-1 and 2.6-2: relating entity B or relation entity R is marked DELETE, R should be updated to remove B from its *RelatedObjects*. Situations 2.7 and 2.8: relation entity R is modified in the sub-model, and simply apply the existence judgement.

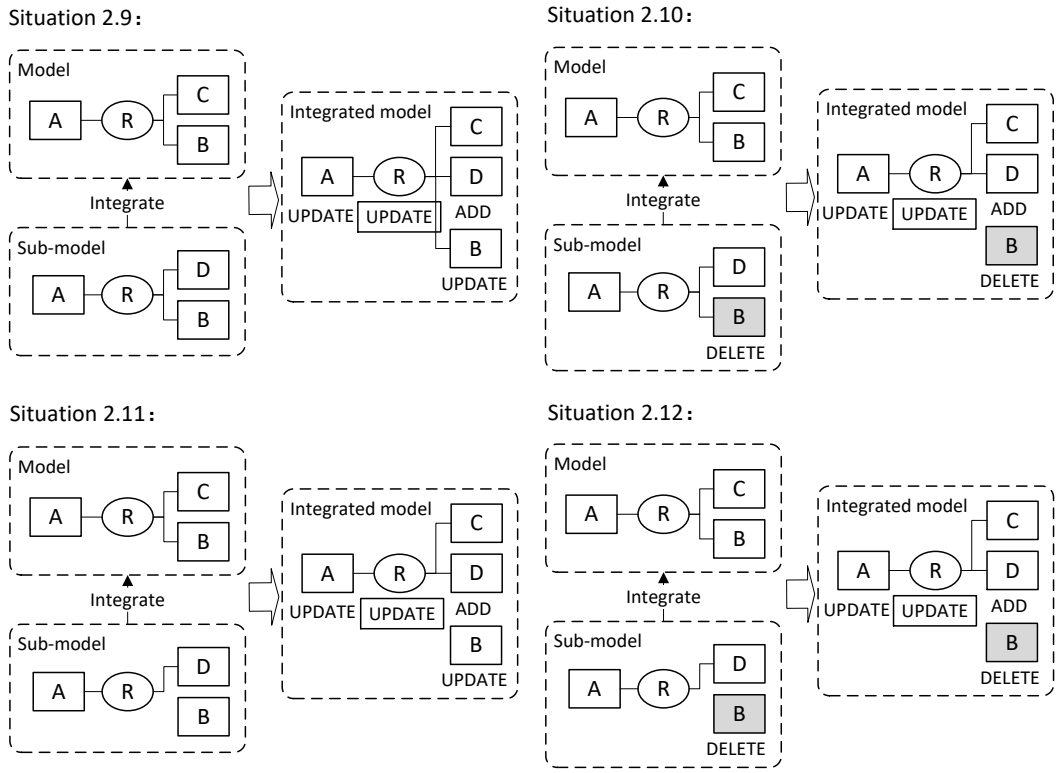

**Figure 14 Integrating situations 2.9-2.12 of [1:n] relationships**

Situation 2.9: no entity needs to DELETE, merge the *RelatedObjects* of relation entity R. Situations 2.10, 2.11 and 2.12: related entity B is marked DELETE or removed from *RelatedObjects*, and it should also be removed from *RelatedObjects* of R in the integrated model.

The states in the black boxes in the above figures are those that cannot be assessed directly from the state analysis, which means that they require corrections. Seven rules are obtained from those situations:

**Rule** 1: if the relating entity or the related entity is deleted in the sub-model, then the [1:1] relation entity itself must be deleted.

**Rule** 2: if the relating entity is deleted in the sub-model, then the [1:n] relation entity itself must be deleted.

**Rule** 3: if the relating entity or the related entity are both in the sub-model but the [1:1] relation entity is not, then the relation entity must be deleted.

**Rule** 4: if a [1:n] relation entity is deleted in the sub-model and not all the related entities are in the sub-model, then it needs to update the relation entity to remove the related entities in the sub-model from the *RelatedObjects* attribute.

**Rule** 5: if a related entity is deleted in the sub-model, then it needs to update the [1:n] relation entity to remove the deleted entity from the *RelatedObjects* attribute.

**Rule** 6: if a [1:n] relation entity has a new related entity in the sub-model, then the new entity should be added to the *RelatedObjects* attribute of the relation entity.

Finally, an additional rule is proposed to avoid null pointers of the [1:n] relation.

**Rule** 7: if all the related entities of a [1:n] relation entity are marked DELETE, or it has no related entities in the integrated model, then the [1:n] relation entity itself should be deleted.

## 4) Overall sub-model integration process

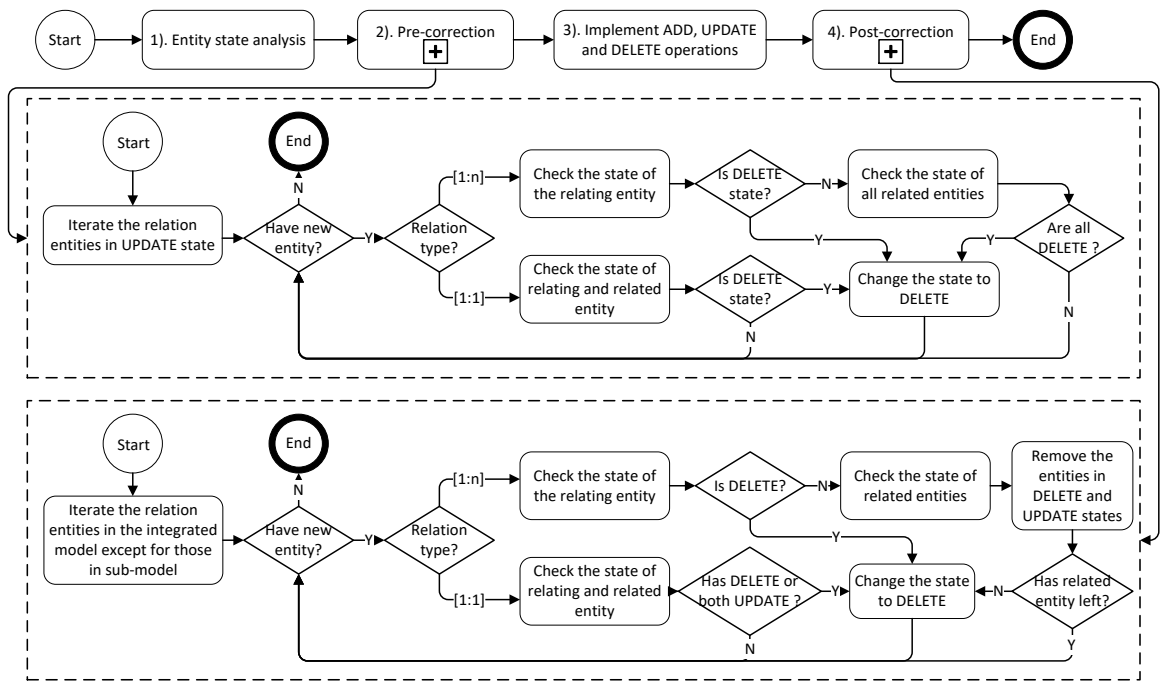

Figure 15 Overall process of sub-model integration

The overall process of the proposed sub-model integration method is shown in Figure 15, which includes four major steps: 1). Analyze the states of the entities in the sub-model, using existence judgement and the DELETE mark; 2). Pre-correction of relationships in the sub-model, applying rules 1, 2 and 5, 6; 3). Implement the ADD, UPDATE and DELETE operations to generate the integrated model; 4). Post-correction of the relationships in the integrated model, applying rules 1, 2, 3, 4, 7.

## 5. Case study and discussion

### 5.1. Prototype system

Based on lessons learnt for BIM tools development from Tsai et al.[19], a prototype system (BIMDISP) was developed together by developers and potential users of the system to establish the presented multi-server data-sharing environment based on BIMIIP, an earlier BIM server developed by Zhang[34] and Yu[7]. The former server performs such functions as IFC parsing, model viewing, MVD defining, and model integration into an IFC database based on an SQL server. The prototype system developed in this study was restructured and designed to satisfy the cross-party use cases and provide services to end users through web browsers and other client applications[7]. The Hadoop HBase was adopted for the underlying data storage. Communication modules and services were developed using WCF and C# .net, and Thrift was used to support cross-language services.

### 1) System architecture

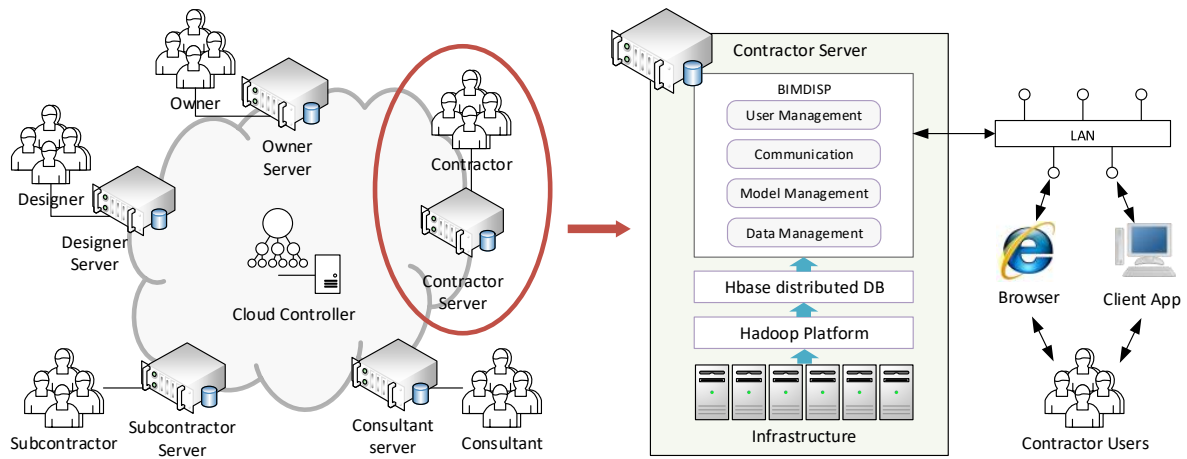

**Figure 16 Architecture of the multi-server BIM data-sharing environment**

Because AEC/FM industry practitioners collaborate with each other in two aspects—collaboration within a party and collaboration between different parties—the multi-server approach fulfills the requirement for data exchange and sharing for both aspects.

According to the designed system architecture shown in Figure 16, a cluster of computers is set up as a server, which is responsible for the multi-disciplinary/multi-user collaboration within a party. Considering the needs for handling large amounts of model data and supporting the potential demand for data mining, the architecture adopts the Hadoop framework for distributed systems and HBase for the storage of the IFC data. In this manner, the storage capacity of each party's server can be easily scaled horizontally by appending the data nodes according to practical needs. For collaboration between different parties, a cloud platform is designed to handle the data exchange and sharing within the distributed multi-server environment. In this environment, the BIMDISP is also responsible for communications among the cross-party servers. These servers are coordinated and scheduled with the help of a cloud controller, as noted previously.

## 5.2. Case study

The developed prototype system of the multi-server data-sharing environment was tested on the data from an actual project of the Kunming new airport terminal to evaluate the following aspects: (1) whether the environment can be successfully and conveniently established using the proposed methods, (2) whether the environment can provide the data in accordance with the actual requirements, and (3) whether the environment performs effectively with large data models.

The Kunming new airport terminal is located in Guandu District of Kunming, the capital city of Yunnan province in China. The construction began in 2008, with a total construction area of 435,400 square meters. The test procedure is as follows:

1) According to the actual needs of the construction process, the party servers were established in a laboratory environment for the owner, MEP general contractor, weak systems contractor, baggage systems contractor, BIM consulting team, and operation and maintenance firm. The party servers were composed of commodity computers. The number of nodes for each cluster is shown in Table 1. A greater number of nodes were configured when a party had larger storage requirements. Each node was a Linux virtual machine with 40 G of hard drive space, 2G of RAM, and two CPU threads.
2) After the system installation and deployment were completed, the IFC databases on each server were automatically created by entering the IFC schema, party MVD files, and server names. The party MVDs were defined using a semi-automated tool developed by Yu[8] according to the data requirements of each party, as shown in Table 1. The IFC2x3 versioned schema was chosen for this test.
3) The test used a self-developed desktop application based on the BIMIIP server as the client and uploaded the model data in the IFC format separately on each party server. The column "produced data" in Table 1 specifies that the data set should be uploaded for each server. As the party MVDs were uploaded, the environment would build and send data copies to form the party sub-models automatically. The volumes of data in each server after completing the upload process are also listed in the table.

**Table 1 Data distribution for multi-parties**

| Party | Nodes | Produced data | Required data | Volume (G) |
|---|---|---|---|---|
| Owner | 5 | | All design and construction models | 98.3 |
| MEP general contractor | 5 | Construction information of the lighting system, binnacles, and corridor ceilings | Design and construction models of interior drainage system and check-in islands; design models of the lighting system, binnacle, and corridor ceilings | 82.1 |
| Weak systems contractor | 3 | Design and construction properties of intelligent-building equipment rooms | Design and construction models of check-in islands, binnacles, and corridor ceilings | 8.5 |
| Baggage systems contractor | 3 | | Design and construction models of check-in islands | 3.5 |
| BIM consulting team | 5 | All design models, construction information on the interior drainage system and check-in islands; O&M models of MEP systems | | 53.4 |
| O&M firm | 10 | Remaining O&M models | All design and construction models | 131.5 |

4) After the data distribution processes, all the parties obtain their required data on their own servers. Figure 17 shows the model view of the weak systems contractor as an example of the distribution result. This figure illustrates that the design and construction models of the check-in islands were obtained from the BIM consulting team, the design model of the binnacles were from the BIM consulting team, and the construction information on the binnacles were from the MEP general contractor, in accordance with the data requirement of the weak systems contractor, as listed in Table 1.

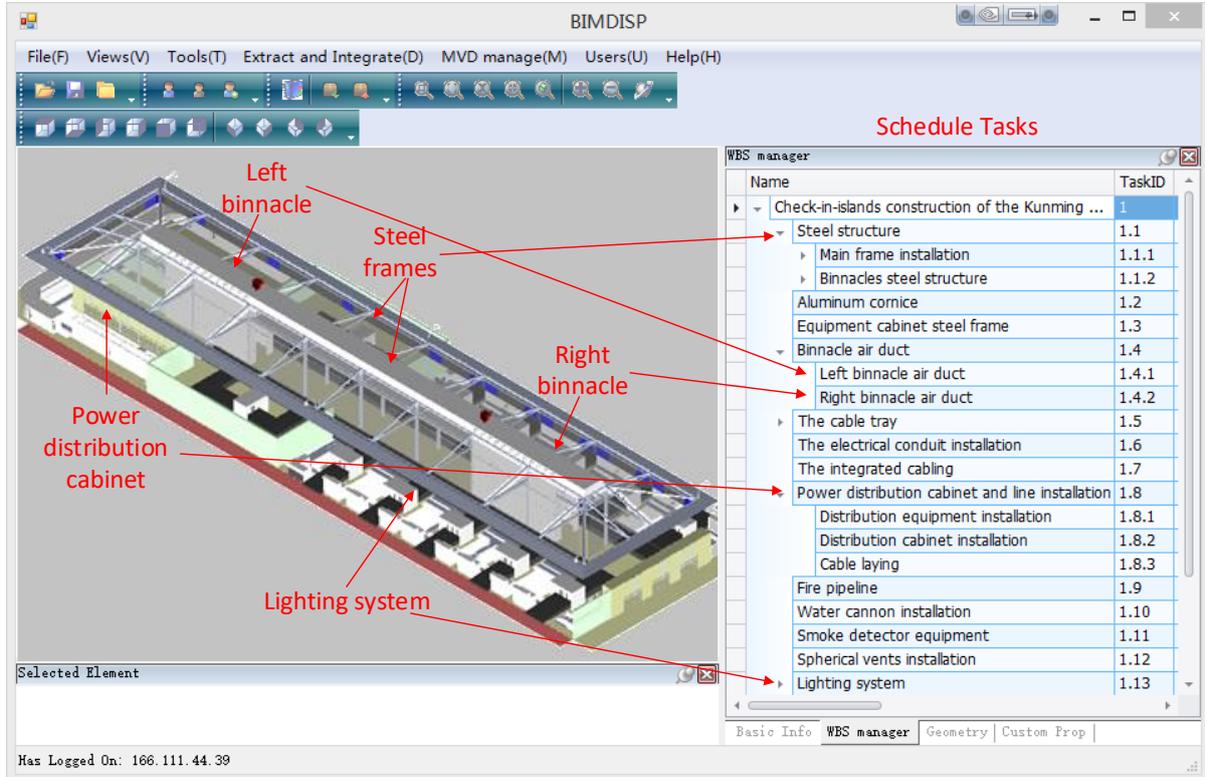

**Figure 17 Model view of the check-in island from the weak systems contractor**

5) During the data distribution processes, the cloud controller builds indexes for all shared data. Table 2 lists the amount of indexing entities in the cloud controller of each party. As mentioned above, data in the server of one party can be categorized by ownership into internal data owned by the party and external data shared by others, and in the internal data, only the shared parts are indexed.

**Table 2 Indexing amount of multi-parties**

| Party | Shared entities | External entities |
| --- | --- | --- |
| Owner | 0 | 2684609 |
| MEP general contractor | 547438 | 2283040 |
| Weak systems contractor | 34831 | 42277 |
| Baggage systems contractor | 0 | 27288 |
| BIM consulting team | 2615230 | 0 |
| O&M firm | 1778356 | 2684609 |

6) A cross-party data extraction process was conducted in this test with the objective of extracting the model that contains the design and construction information of the interior drainage system in Region A on floors B2 and B3. The required data are produced by both the BIM consulting team and MEP general contractor. The MVD for the extraction was defined, containing 79 types of entities inherited from the IfcRoot entity, among which 20 were type entities and 24 were relation entities, as well as 127 types of entities from the resource layer. The received IFC data included 2,016,579 objects with a size of 121.764 m in the form of an "*.ifc" file and included a "B3 A interior drainage system" and "B2 A interior drainage system," two root tasks and 54 sub-tasks, and a total of 6,878 physical components of the B3 A and B2 A district interior drainage systems. The extraction results in the client application, as shown in Figure 18, are in accordance with expectations: The required components, spatial structure, schedule, and other relevant data were successfully extracted from the data producers and integrated into a sub-model as a result of the request.

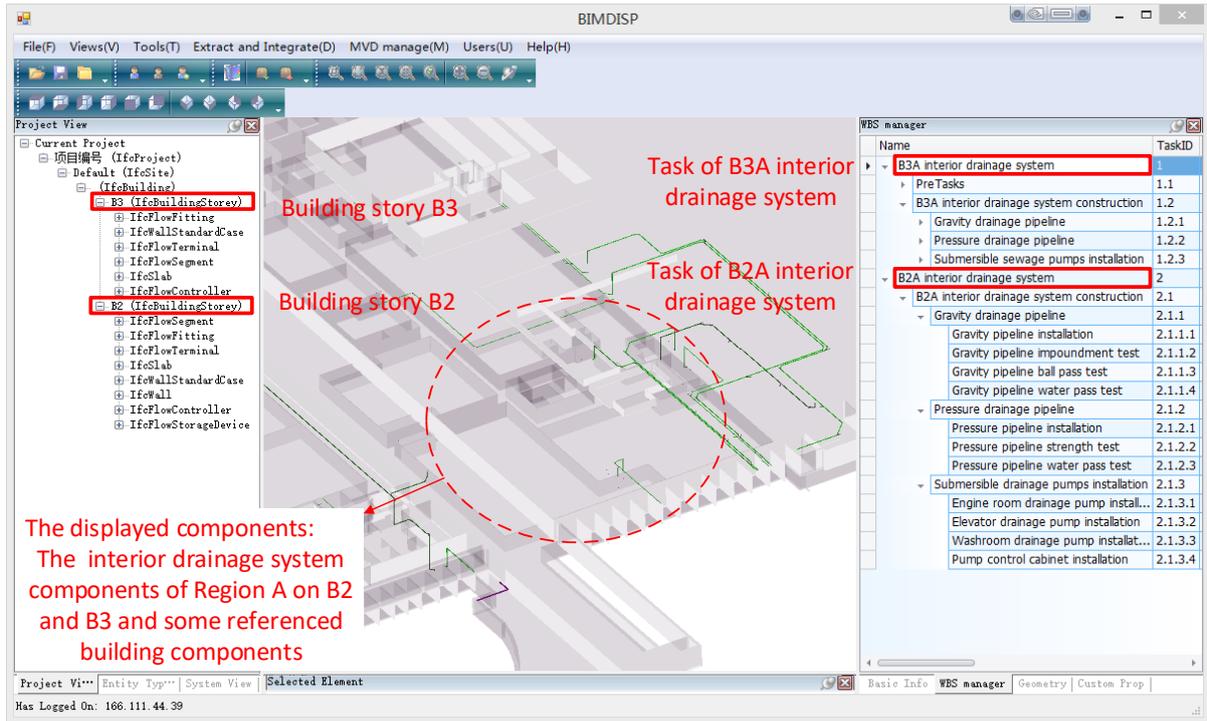

Figure 18 Extraction results from the test on the new Kunming airport terminal

7) In order to verify the performance of the proposed parallel and distributed data extraction method, a model extraction test was conducted. The test used BIMIIP and Yu's model extraction method[7] as the control group, the results are shown in Table 3. The comparison shows that the extracted sub-model of the improved method is basically the same as that of the existing method. The comparison of the extraction time indicates that the improved method can extract the sub-model from the database faster, especially for complex tasks.

Table 3 Results of model extraction test

| Extraction condition | Existing method | | Improved Method | |
|---|---|---|---|---|
| | Major entities | Time (s) | Major entities | Time (s) |
| IfcTask.Name= "Steel structure" | 1 IfcProject, 4 IfcSpatialStructure, 5 IfcProcess, 188 IfcProduct | 18.66 | 1 IfcProject, 4 IfcSpatialStructure, 5 IfcProcess, 188 IfcProduct | 12.48 |
| "Binnacle main structure"∈IfcBuildingElementType.Name | 1 IfcProject, 3 IfcSpatialStructure, 1 IfcProcess, 20 | 10.43 | 1 IfcProject, 3 IfcSpatialStructure, 20 IfcProduct | 9.52 |
| ∩ IfcBuildingStorey.Name ="Ground" | IfcProduct | | | |
| IfcActor.Name= "MEP general contractor" ∩ IfcTask.Name= "Steel structure" | 1 IfcProject, 1 IfcOrganization, 3 IfcSpatialStructure, 21 IfcProcess, 374 IfcProduct | 34.65 | 1 IfcProject, 1 IfcOrganization, 3 IfcSpatialStructure, 21 IfcProcess, 374 IfcProduct | 14.51 |

### 5.3. Discussion

a) The main obstacle to verifying the application of the approach is the lack of commercial tools, which should be IFC-compliant and be able to ensure data consistency in a complete import–process–export process; thus, only the tools developed by our group were tested.

b) Although the study used Hadoop HBase and WCF to develop the single-server modules of each party in the prototype system, the database and development platform were not limited for the multi-server approach and environment. This paper mainly presents the general methods and design principles. Various technical means can be used in the implementation.

c) Compared with the centralized approaches, the multi-server approach is more complex and introduces data redundancy. However, the increasing complexity provides significant benefits that guarantee the usability of the data for the user. Different from the file exchange approach, the redundancy can be controlled in the multi-server environment. In addition, the multi-server approach allows parties to fully control the software and hardware of their party servers, thus providing more favorable conditions to protect private data compared to the centralized approaches.

d) Some assumptions in the conventions for the consistency issue noted in Section 4.1 apply only to cross-party scenarios. For example, data ownership is commonly exclusive to an enterprise, but data are unlikely to be exclusive in multi-disciplinary or multi-user collaboration scenarios because many people must modify the same data in these situations. In other words, the multi-server architecture provides a distinction between the two levels of scenarios. Thus, the consistency management strategy can be developed separately according to different requirements.

e) In actual applications, deploying the cloud controller can be controversial. The principle of the cloud controller can be a threat to data privacy, although technical means are applied to address this issue. Policy and legal efforts are necessary in this situation. Currently, the most reasonable proposal is to deploy the controller on the server of the project owner.

f) Introducing a new collaboration always brings organizational changes to project teams and associate stakeholders, possible framework should be adopted to manage these changes and their impacts on project success[4].

### 6. Conclusions

By considering the features of the BIM data and the special requirements, such as data ownership and privacy in cross-party collaboration, this study proposes a multi-server approach based on a private cloud for BIM data sharing and management among different stakeholders. In this approach, the multi-party users are allowed to store and access the relevant data in their own servers, and the overall data are distributed in the composed private cloud platform. Global data controlling and authorization, distributed sub-model extraction and integration in a distributed environment are investigated in depth. A global controller was proposed to register and track the data distribution and authorization based on the predefined consistency conventions. Furthermore, parallel data processing was introduced in sub-model extraction to fully utilize the power of a distributed platform. Additionally, entity state analysis and relationship correction were used in sub-model integration to avoid consistency issues. Finally, a prototype system that adopts the proposed methods and approaches was developed to evaluate the proposed approach. In summary, the multi-server data-sharing environment is advantageous in terms of data ownership, privacy control and consistency maintenance and in terms of massive data storage, transfer and processing. With the proposed approach, different stakeholders can collaborate with one another smoothly while retaining ownership and control of their data.

**Acknowledgments**

This work was supported by the National Natural Science Foundation of China (No. 51278274, No. 51478249), the Beijing Municipal Science and Technology Project (No. Z151100002115054), and the Tsinghua University-Glodon Joint Research Center for Building Information Model (RCBIM). The authors also thank Kincho H. Law (Stanford University) for his valuable comments and contributions.